\documentclass[12pt]{iopart}

\usepackage{color}
\usepackage{epsf}
\usepackage{graphicx}
\usepackage{iopams}  
\begin{document}

\title [Spin-fluctuations in Ti$_{0.6}$V$_{0.4}$] {Spin-fluctuations in Ti$_{0.6}$V$_{0.4}$ alloy and its influence on the superconductivity}

\author{Md Matin, L S Sharath Chandra, Radhakishan Meena, M K Chattopadhyay, S B Roy}
\address{Magnetic and Superconducting Materials Section, Raja Ramanna Center for Advanced Technology, Indore, Madhya Pradesh - 452 013, India}
\author{M N Singh, A K Sinha}
\address{Indus Synchrotrons Utilization Division, Raja Ramanna Center for Advanced Technology, Indore, Madhya Pradesh - 452 013, India}
\ead{lsschandra@rrcat.gov.in}

\begin{abstract}  
We report experimental studies of the temperature and magnetic field dependence of resistivity and dc magnetic susceptibility and the temperature dependence of zero field heat capacity in a Ti$_{0.6}$V$_{0.4}$ alloy. The temperature dependence of the normal state dc magnetic susceptibility in this Ti$_{0.6}$V$_{0.4}$ alloy shows T$^2$lnT behavior. The temperature dependence of resistivity follows a T$^2$ behaviour in the range 20-50~K. On the other hand, a term $T^3$ lnT is needed in the expression containing the electronic and lattice heat capacities to explain the temperature dependence of heat capacity at temperatures where $T^2$ dependence of resistivity is observed. Such temperature dependence of dc magnetic susceptibility, resistivity and heat capacity are indications of the presence of spin-fluctuations in the system. Further experimental evidence for the spin fluctuations is obtained in the form of a negative value of T$^5$ term in the temperature dependence of resistivity. The influence of spin-fluctuations on the superconducting properties of Ti$_{0.6}$V$_{0.4}$ is discussed in detail. We show from our analysis of resistivity and the susceptibility in normal and superconducting states that the spin fluctuations present in Ti$_{0.6}$V$_{0.4}$ alloy are itinerant in nature. There is some evidence of the existence of preformed Cooper-pairs in the temperature range well above the superconducting transition temperature. Our study indicates that the interesting correlations between spin-fluctuations and superconductivity may actually be quite widespread amongst the superconducting materials, and not necessarily be confined only to certain classes of exotic compounds.  
\end{abstract}

\maketitle

\section {Introduction}

The role of spin-fluctuations in superconducting materials has been a subject of sustained interest for the last 50 years. In 1960s Doniach \cite{don67} and Berk and Schrieffer \cite{ber66} made the important observation that itinerant ferromagnetic spin-fluctuations tend to suppress superconductivity in elemental s-wave superconductors. They pointed out that the absence of superconductivity in heavy transition elements such as Palladium and Platinum is due to the itinerant ferromagnetic spin-fluctuations resulting from the large $d$-electron density of states near the Fermi level. The same formalism was used to suggest that the superconducting transition temperature T$_c$ of elements such as Nb and V was limited by spin-fluctuations \cite{rie79, wei05}. Hence it was conjectured that when these metals were properly engineered in the form of alloys, the T$_c$ could increase significantly as was observed in Nb$_3$Sn, and V$_3$Si, etc., \cite{mit84, mit86}. Later on, the effects of localized spin-fluctuations on the properties of dilute superconducting alloys were studied by Zukermann \cite{zuc70}. He showed that the superconducting transition shifted to  temperatures lower than those expected from the band structure calculations performed without considering the spin-fluctuations \cite{zuc70}. On the other hand, many families of superconductors have been identified during last 30 years, where spin-fluctuations actually play an important role in the superconductivity itself. These include various exotic low temperature superconductors such as Y$_9$Co$_7$ \cite{sar86}, CeCoIn$_5$ \cite{kaw03}, UBe$_{13}$ \cite{hef91}, Sr$_2$RuO$_4$ \cite{kuw00}, FeSe \cite{qiu09}, and Mo$_3$Sb$_7$ \cite{can07} etc. This concept of spin-fluctuation influenced superconductivity, however, came into much prominence in connection with the high temperature superconductors \cite{jin11, dah09, tac11}. This has now become a subject of even more attention with the discovery of a newer class of Fe based high temperature superconductors \cite{kam08, hir11}. Most recently it has been reported that the longitudinal ferromagnetic spin-fluctuations induce superconductivity in UCoGe \cite{hat12}.

It is well known that any addition of a magnetic impurity in non-transition element based s-wave superconductor suppresses the superconductivity due to the pair breaking. It is also reported in literature that even the addition of the non magnetic transition elements also suppresses the T$_{c}$ \cite{von82, aok67, aok69} due to the formation of localized states. However, it has been observed that the T$_c$ of a dirty limit superconductor is not effected significantly by disorder \cite{and59}, in fact a very high level of disorder is required to change the T$_c$ \cite{sac11}. Surprisingly an enhancement in T$_{c}$ is observed in spite of increased disorder when certain transition elements are alloyed, even though these elements either have T$_c$'s lower than the host material, or are non superconducting \cite{von82}. The present Ti-V alloys are the examples of one such system \cite{von82, col75}. In recent times, it has been recognized that Ti-V alloys with their good mechanical, thermal and superconducting properties \cite{col86} can be promising candidates for application as superconducting magnets in the environment of heavy neutron irradiation as in a fusion reactor \cite{tai07}. In addition, higher value of lower-critical field H$_{c1}$(0) of Ti-V alloys at 2~K \cite{bla65} as compared to the commercial polycrystalline Nb samples \cite{roy08} makes them potential candidates for superconducting RF cavity applications. Thus a complete understanding of the normal and superconducting states in these Ti-V alloys can lead to the possibility of enhancing their superconducting properties (e.g., increasing the experimental superconducting transition temperature T$_c$, the lower critical field H$_{c1}$(0) and the upper critical field H$_{c2}$(0) to the theoretical limits) and enhancing other functional properties like critical current density by suitable engineering techniques.     

However, several features of Ti-V alloys are not understood yet, of which we list below the two significant anomalies.    

1. There is an increase in T$_c$ when Ti content is increased from zero to 40 at. \% in V \cite{col75}. The significant enhancement in Debye temperature $\theta_D$ or the electron phonon coupling is not expected in Ti-V alloys as Ti and V are adjacent atoms. The density of states decreases as Ti content is increased in V \cite{col75, von82}. Hence, the increase in T$_c$ in Ti-V alloys does not follow the McMillan formula. 

2. There is a large region in H-T phase diagram where superconducting fluctuations are observed well above T$_c$ and well above H$_{c2}$ \cite{hak69, hak70, lue75}. Hake and coworkers \cite{hak69, hak70, lue75} have shown that superconducting fluctuations well above T$_c$ and well above H$_{c2}$ are intrinsic property of certain class of transition element binary alloys such as Ti-V, Ti-Mo and Ti-Ru systems. They have also observed that these fluctuations are not dependent on the sample preparation, surface polishing, size and shape of the sample and on the current density.

Several models such as the occurrence of reversible $\omega$ phase in the $\beta$ matrix \cite{col78, col86, sas87}, weak localization \cite{sas90}, Kondo (s-d) interaction \cite{pre73} and associated localized spin fluctuation \cite{ras74, pre73, pre75, pre75a} have been considered for explaining the observed physical properties in Ti-V alloys. The increase in magnetic susceptibility with temperature was attributed mainly to the occurrence of reversible $\omega$ phase in the $\beta$ matrix \cite{col78}. According to this model, the metastable hexagonal closed pack (hcp) $\omega$ phase appears inside the body centered cubic (bcc) $\beta$ phase when the temperature is decreased below 300~K. However, this model does not explain the large difference between the experimental value of superconducting transition temperature T$_c$ and that estimated theoretically \cite{pic87}. Prekul et al \cite{pre75} on the other hand invoked the idea of the localized spin fluctuations to understand the normal state properties of Ti-V alloys. However, this model could not explain the superconductivity fluctuations observed well above T$_c$ and well above H$_{c2}$ \cite{pre75a}. 

In this direction we present a detailed study of the superconducting and normal state properties of a Ti$_{0.6}$V$_{0.4}$ alloy investigated through the measurements of temperature dependence of electrical resistivity, dc magnetic susceptibility and heat capacity in the presence of zero and externally applied magnetic fields. The observed temperature dependence of the normal state properties is explained within the realm of itinerant spin-fluctuation model. 
   
\section {Experimental}
The sample was prepared by taking high purity Titanium (99.99~\%, Alfa Aesar) and Vanadium (99+~\%, Aldrich) in stoichiometric proportion and melting in an arc furnace under high purity argon atmosphere. The sample was flipped and re-melted four times to ensure homogeneity. The sample was then wrapped in Ta-foil and sealed in quartz ampoule in argon atmosphere, and was annealed at 1573~K for 10 hours. The sample temperature was then lowered to 1273~K before quenching into the ice water. The angle dispersive X-ray diffraction (XRD) measurement were performed in a powder XRD beam line BL-12 \cite{sin11} at 19~keV (0.65 x 10$^{-10}$~m) X-rays from INDUS-2 synchrotron radiation source at Raja Ramanna Centre for Advanced Technology, Indore. The beam line is based on Si (311) double crystal monochromator and adaptive focusing optics.  An image plate area detector (MAR-345 dtb) and Fit2d software were used for data acquisition and data reduction, respectively. NIST LaB6 standard was used for wavelength calibration. The resistivity measurements as functions of temperature and magnetic field were performed in a 5~T magnet-cryostat (American Magnetics Inc, USA) using a standard four-probe geometry. The magnetization measurements were performed using a 9~T Vibrating Sample Magnetometer (VSM; Quantum Design, USA) and a 7~T SQUID magnetometer (MPMS-XL; Quantum Design, USA). The heat capacity measurements were performed using a 9~T Physical Property Measurement System (PPMS; Quantum Design, USA).

 \section {Results and Discussion}
 \subsection{Structural characterization of Ti$_{0.6}$V$_{0.4}$ using x-ray diffraction}


The Ti-V alloys form in $\beta$ phase for Ti composition exceeding 20 at. \%. The space group corresponding to $\beta$ phase is body centered cubic (bcc) Im$\bar{3}$m with Ti or V occupying 2a site randomly. Figure 1 shows the XRD pattern of Ti$_{0.6}$V$_{0.4}$ alloy along with the fitted curve obtained using Rietveld refinement. All the peaks observed in XRD pattern are correspond to the $\beta$ phase in the Ti$_{0.6}$V$_{0.4}$ alloy.  The Rietveld analysis (red solid line) shows that the sample is formed in bcc structure with lattice parameters 0.31879(2)~nm, which is in agreement with the literature \cite{aur02}.

\subsection{Superconducting properties of Ti$_{0.6}$V$_{0.4}$}






Figure 2 presents the superconducting properties of Ti$_{0.6}$V$_{0.4}$. Figure 2a shows the temperature dependence of magnetization measured using VSM in the temperature range 2-7.2~K in an applied field of 10~mT. The measurements were performed in the zero field cooled (ZFC), field cooled cooling (FCC) and field cooled warming (FCW) modes. In the ZFC mode, the sample is first cooled from a temperature well above 10~K to 2~K in zero field and then the measurements are performed while warming up the sample in presence of 10~mT. In the FCC mode, the magnetic field (10~mT) is applied at a temperature well above 10~K, and the measurement is done while cooling the sample down to 2~K. After reaching 2~K, the measurement is continued while warming up the sample in the same field and this last protocol is called FCW. The superconducting transition temperature T$_c$ is the temperature at which the M(T) starts to drop towards the negative value when temperature is decreased. The T$_c$ thus estimated at 10~mT is about 7.15~K, which is in agreement with the literature \cite{col75}. The T$_c$ of Ti$_{0.6}$V$_{0.4}$ is higher than that for elemental Vanadium (5.4 K \cite{col75, von82}) or Titanium (0.9~K \cite{col75, von82}). The Meissner fraction is estimated as M$_{FCC}$(2 K)/M$_{ZFC}$(2K) is about 7 x 10$^{-4}$ at 10~mT which indicates the strong flux pinning. Figure 2b shows the magnetization as a function of magnetic field measured using VSM at various constant temperatures below superconducting transition temperature 7.15~K. The measurements were performed as a function of magnetic field after cooling the sample from well above the 10~K to the desired temperature in zero magnetic field. There is a considerable hysteresis at low fields in all the isothermal M-H curves. For field greater than a characteristic field H$_{irr}$ magnetization becomes reversible. The upper inset (i) of the Fig. 2b shows the M v/s H data near H$_{c2}$ and the lower inset (ii) shows the low field M v/s H data at selected temperatures which are used to estimate the H$_{c1}$ from the magnetization data. We plot in Figure 2c the temperature dependence of resistivity below 15~K at various constant magnetic fields up to 5~T. In zero magnetic field, the resistivity decreases as temperature is lowered below 15~K and drops to zero at T$_c$ = 7.06~K with a transition width of 0.085~K. The inset in Fig. 2c shows the effect of magnetic field (up to 5~T) on the superconducting transition temperature; an application of 5~T suppresses the T$_c$ to 5.66~K. We have observed a rounding-off behavior in the temperature dependence of resistivity just above the T$_c$. Figure 2d shows the temperature dependence of heat capacity plotted as C/T as a function of T$^2$ in selected magnetic fields. The application of magnetic field shifts the heat capacity peak to lower temperature. The superconducting transition temperature T$_{c, o}$ is obtained as that temperature at which a deviation is observed from the normal state linearity in C/T v/s T$^2$ plot. In zero magnetic field, T$_{c, o}$ is estimated to be about 7.17~K. At 8~T magnetic field, the superconducting transition is observed at 4.8~K.There is no difference in normal state heat capacity between the zero field data to that in magnetic fields. The Sommerfeld coefficient of electronic heat capacity $\gamma$ and the Debye temperature $\theta_D$ are obtained to be about 9.46(2) x 10$^{-3}$ J/ mol K$^2$ and 258.5(2) K respectively by fitting a straight line to the normal state heat capacity data in figure 2d. The jump in heat capacity $\Delta$C/$\gamma$T$_c$ across the T$_c$ in the zero magnetic field is estimated at the onset of the transition T$_{c, o}$ as well as at the middle of the transition T$_{c, m}$ = 6.63~K. The $\Delta$C/$\gamma$T$_{c, o}$ is about 2.65 whereas $\Delta$C/$\gamma$T$_{c, m}$ is about 2.22 indicating that Ti$_{0.6}$V$_{0.4}$ is a bulk and a strong coupling superconductor. The value of T$_{c, o}$ is in agreement with the T$_c$ estimated using resistivity and magnetization measurements.


The upper critical field H$_{c2}$ at various temperatures were estimated from such isothermal M-H curves is shown in Fig. 3a. The magnetic field at which a distinct deviation takes place from the magnetic field dependence of the normal state magnetization is taken as the H$_{c2}$. A straight line is fitted to the M v/s H data for the normal state M just above the superconducting state. The H$_{c2}$ is taken as that point at which the difference between the experimental data and the fitted line exceeds the standard deviation of the fitting. This procedure has been effectively used to estimate the H$_{c2}$ in superconductors such as borocarbides \cite{roy94, roy96} and skutterudite \cite{sha12} where enhanced paramagnetism is observed in the normal state. The error in the estimation of H$_{c2}$ is found to be less than 2~\%. The temperature dependence of H$_{c2}$ is analyzed on the basis of formalism by Orlando et al \cite{orl79} (red solid line) which will be discussed in detail later. The H$_{c2}$(T = 0) is estimated to be about 13.68~T. The lower critical field H$_{c1}$ (Fig. 3b.) is estimated as that field where the slope of the linear fit to the low field data (inset (ii) to Fig. 2b.) deviates about 2~\% from the value of 1. The magnetization data at low fields are corrected for demagnetization effects before estimating the H$_{c1}$. The T$_c$ estimated from the experimental H$_{c1}$(T) is about 7.1~K. The temperature dependence of H$_{c1}$ follows a T$^2$ dependence. The experimental data deviates from the T$^2$ fit near T$_{c}$. The H$_{c1}$(0) is estimated by the T$^2$ fit to be about 24.38(1)~mT.




The rounding-off behavior of electrical resistivity shown in Fig. 2c. is an indication of fluctuation conductivity above the T$_c$ (7.06~K). In order to verify this, we have studied this region through the estimation of excess conductivity $\Delta \sigma $ = $\sigma_{exp} - \sigma_n$ in zero magnetic field (Fig. 4a.) and the magneto resistance $\Delta \rho/\rho$ = ($\rho$(H) - $\rho$(0)) x 100 / $\rho$(0) as a function of magnetic field at temperatures above the T$_c$ (Fig. 4b.). Figure 4a shows the log-log plot of $\Delta \sigma $  as a function of (T-T$_c$)/T$_c$ in the temperature range 7.15-15~K. The excess conductivity may be explained in terms of the additional conductivity that results from the formation of Cooper pair pockets well above T$_c$; these Cooper pairs are yet to be in the condensed state. The $\sigma_n$ stands for the normal state conductivity, which is generally estimated by extrapolating the temperature dependence of resistivity from a temperature well above 3 times of T$_c$ to the region where the fluctuation conductivity is observed \cite{gau95}. The $\sigma_{exp}$ is the experimentally observed conductivity, and the transition temperature T$_c$  = 7.06~K is taken as the temperature at which the temperature-derivative of resistivity shows a maximum.

According to theory of Aslamazov and Larkin (AL) \cite{asl68, jho76}, the excess conductivity $\Delta \sigma_{AL}$ is expressed as  

\begin{equation}	
\Delta \sigma_{AL} =\sigma_{exp} - \sigma_n = At^{-\alpha} .
\end{equation}

where, $t$=(T-T$_c$)/T$_c$ is the reduced temperature, $A$ is a constant (temperature independent), $\alpha$=2-$D$/2 is the critical exponent, and $D$ is the dimensionality of the superconducting fluctuations. For a three dimensional superconductor, $A = e^2/32\hbar \xi(0)$ and $\alpha$= 0.5, where $\xi$(0) is the coherence length at zero temperature \cite{asl68, jho76}. The straight line fit near T$_c$ to the data in Fig.4a., shows that  $\alpha$= 0.5 indicating 3$D$ character of the superconducting fluctuations in Ti$_{0.6}$V$_{0.4}$. However, the temperature range over which this fluctuation is observed is limited to 7.25 to 8.03~K. The upturn in $\Delta \sigma $ below 7.25~K is due to the broadening of the superconducting transition \cite{gau95}. The Maki - Thompson (MT) \cite{mak68, mak68a, tho70, jho76} effect can also contribute to the fluctuation conductivity where electron -electron interactions are strong. However, the $\xi$(0) estimated from the temperature independent amplitude of the AL contribution to fluctuation conductivity is about 4.49~nm which is in agreement with the $\xi$(0) estimated from H$_{c2}$(0). This indicates that the MT contribution to fluctuation conductivity is not dominant in Ti$_{0.6}$V$_{0.4}$. Apart from this, MT contribution is dominant only in the clean limit \cite{jho76} whereas Ti$_{0.6}$V$_{0.4}$ is in the dirty limit. It may be noted here that the AL model is valid only in limited range of temperature close to T$_c$. The figure 4b shows $\Delta \rho/\rho$ as a function of magnetic field at temperatures above the T$_c$. At 7.5~K and 5~T, the $\Delta \rho/\rho$ is about 0.45~\%. The $\Delta \rho/\rho$ at 5~T decreases as the temperature increases. Hence, from Fig. 4b, it appears that the magneto-resistance in Ti$_{0.6}$V$_{0.4}$ exists up to 3T$_c$, where as Fig. 4a., shows that fluctuation conductivity exists up to at least 2T$_c$. Such positive magneto-resistance below 3T$_c$, which is linear in magnetic field, was interpreted by Hake to be due to the breaking of these pre-formed Cooper pairs which formed well above T$_c$ but were not condensed yet to give rise to  superconductivity \cite{hak69, hak70}. In this latter argument, the temperature and magnetic field dependence of resistivity in the temperature range T$_c$=7.15~K to 14.3~K in Ti$_{0.6}$V$_{0.4}$ is due to the excess conductivity or the fluctuation conductivity resulting from Cooper pair fluctuations well above T$_c$. Then the effect of the magnetic field is to break the Cooper pairs and drive the sample to its normal state. Lue et al have shown experimentally that the magnetic field required to drive a system into normal state is more than the H$_{c2}$(0) \cite{lue75} for temperature even above T$_c$. The observed large positive magnetoresistance above T$_c$ cannot be accounted by the AL fluctuation conductivity in presence of magnetic field for a 3D superconductor in the dirty limit is given by \cite{sam98, the97, ull91}. In case of Niobium as well as alloy and compound superconductors such as Nb$_3$Sn, it is reported that the disorder generated by the high flux neutron irradiation does not decrease the T$_c$ significantly \cite{ker67, sek78}. As pointed out earlier \cite{and59}, the moderate disorder does not influence the transition temperature in dirty limit superconductors. Very high disorders are required to drive the T$_c$ to lower temperatures which are accompanied by a transition from superconducting state to insulating state at T$_c$ \cite{sac11}. Since, our system is a metallic and when Vanadium is alloyed with the Titanium, the T$_c$ of the alloy increases, the origin of the observed fluctuation conductivity in Ti$_{0.6}$V$_{0.4}$ may not be due to the disorder present in the alloy. However, in our alloy, the coherence length is about twice the unit cell. Hence it is necessary to consider the effect of disorder on the superconducting properties. The transition from fluctuation conductivity region to condensate region ($\rho$ = 0) in resistivity for Ti$_{0.6}$V$_{0.4}$ takes place over a temperature width $\Delta T_C$  = 0.085~K centered on T$_c$ = 7.06~K. This transition temperature width $\Delta T_C$ is very large when compared to the $\Delta T_C$ of Nb and V which is about 0.001-0.01~K \cite{smi70, rad66}. The $\Delta T_C$  is estimated by extrapolating the derivative of temperature dependence of resistivity to zero from both side of T$_c$. Within the Landau theory of phase transitions \cite{jho76}, the transition broadening due to disorder is given by $(\Delta T_C^2)^{1/2}$ = (dT$_c$/dx) ($\Delta$x$^2$)$^{1/2}$ where, ($\Delta$x$^2$)$^{1/2}$ is the root mean square (RMS) fluctuation in the composition. For a A$_{1-x}$B$_x$ binary alloy, the average number of A and B atoms are $(1-x)$N and N$x$ respectively, where N = V$_c$/$\nu_q$ is the number of atoms in the characteristic volume V$_c$=(4/3)$\pi \xi$(0)$^3$ and $\nu_q$ is the mean atomic volume. If the alloy is taken to be ideally random, then the RMS deviation in the number of B atom $\sim$(N$x$)$^{1/2}$ and the RMS deviation in the compositional variable, ($\Delta$x$^2$)$^{1/2} \sim$ $x/(Nx)^{1/2}$. For Ti$_{0.6}$V$_{0.4}$, $\nu_q \sim$ 1.108 x 10$^{-29}$~m$^3$, and V$_c$ $\sim$ 4.78 x 10$^{-25}$~m$^3$ which leads to N $\sim$ 3.93 x 10$^4$. The dT$_c$/dx for Ti-V alloys is about 0.05~K/at., for V rich samples \cite{col75}. Then the $\Delta$T$_c$ due to disorder is about 16~mK which is distinctly small as compared to that observed experimentally. The transition broadening due to the fluctuation of atomic density \cite{jho76} is also observed to be negligibly small.  

Hence, we believe that the disorder is not the origin for the observed physical properties viz.,

1. The increase in T$_c$ when Ti content is increased from zero to 40 at. \% in V. 

2. The presence of fluctuation conductivity and superconducting origin of magneto resistance above T$_c$.

In order to understand the origin of these superconducting properties, we have performed a detailed study on the normal state properties of the Ti$_{0.6}$V$_{0.4}$ alloy, which is presented below.

\subsection{Normal state properties of Ti$_{0.6}$V$_{0.4}$}



Figure 5(a) shows the temperature dependence of dc susceptibility $\chi$ of Ti$_{0.6}$V$_{0.4}$ alloy in the temperature range of 10-300~K measured using SQUID magnetometer in an applied magnetic field of 1~T.  The data is corrected for background signal. The dc susceptibility is defined as the ratio of measured magnetization (M) and applied magnetic field (H). As the temperature decreases below 300~K, the $\chi$(T) decreases down to about 30~K before showing an upturn at still lower temperature. The isothermal field dependence of magnetization at 10~K measured with background correction using SQUID magnetometer up to 7~T is shown in Fig. 5(b). The isothermal field dependence of magnetization at various temperatures T = 10, 20, 30, 50, 100, 150, 200, 250 and 300~K is also measured up to 8~T using vibrating sample magnetometer. For the sake of clarity, we have provided such isothermal M v/s H curves at 10, 30 and 300~K in Fig. 5(b). These curves does not show any indications of saturation even at 8~T. These curves rule out any predominant contribution from ferromagnetic impurities. Since, the difference between M v/s H curves at various temperatures is not clearly seen in Fig. 5(b), we have plotted the difference M(H, T) - M(H, 300~K) in the inset to Fig. 5(b) which clearly demonstrate the effect of temperature on the magnetic susceptibility. The susceptibility estimated as the high field slope of the isothermal M v/s H data at various temperatures normalized to that at 30~K is shown in the inset to the figure 5(a). For comparison we have also plotted the temperature dependence of $\chi$(T) at 1~T (measured with SQUID magnetometer) normalized to that at 
30~K.

The increase in the dc susceptibility with temperature has been observed earlier in various 4d and 5d transition metals and rare earth and actinide based paramagnetic intermetallic compounds (Ref. \cite{mis76} and Ref. \cite{bar75} and the references therein) and  Mo$_3$Sb$_7$ \cite{can07}.  

The temperature dependence of susceptibility of such paramagnetic metals is observed to follow \cite{mis76, bar75} 

\begin{equation}	
\chi (T) =\chi (0) -bT^2 ln (T/T^*).
\end{equation}

Here T$^*$ is a characteristic temperature which is related to the peak position T$_{peak}$ in temperature dependence of susceptibility as T$_{peak}$ = T$^*$/$\sqrt{e}$ with $e$ as natural logarithmic base, $b$ is a temperature independent constant. The solid red line in Fig. 5(a) show the fit using above equation (2) to the experimental $\chi$(T) of the Ti$_{0.6}$V$_{0.4}$ alloy along with a Curie-Weiss term for the low temperature upturn. This low temperature feature appears to be due to paramagnetic impurities (such as other transition metal elements, Oxygen, Nitrogen and the oxides and nitrides of Vanadium as well) as in the case of Vanadium \cite{kri54, chi59, gal74, hec76}. The values of the fitting parameters are as follows: $\chi$(0) = 4.92 x 10$^{-10}$~WbA$^{-1}$m$^{-1}$, $b$ = 3.88 x 10$^{-16}$~WbA$^{-1}$m$^{-1}$K$^{-2}$, T$^*$ = 512~K. The errors in the parameters estimated are less than 10~\%.


Figure 6 presents the resistivity of the Ti$_{0.6}$V$_{0.4}$ alloy in the temperature range 5-300~K in zero magnetic field. At 300~K the electrical resistivity is about 105.5 ~$\Omega$m. The electrical resistivity decreases linearly down to 90~K. The decrease in resistivity is sharper when the temperature is lowered below T$^*$ = 90~K.  The residual resistivity $\rho_0$ is about 99 x 10$^{-8}$~$\Omega$m. The resistivity goes to zero below 7~K as the system becomes a superconductor. The inset to the Fig. 6 shows the resistivity of Ti$_{0.6}$V$_{0.4}$ plotted as a function of T$^2$. The resistivity in the temperature range 20-50~K is observed to be linear in T$^2$ and can be expressed as $\rho$ (x 10$^{-8}$~$\Omega$ m) = 98.98 + 5.05 x 10$^{-4}$ T$^2$. The error in the coefficients of linear fit is less than 0.1~\%. However, the plot of ($\rho$-$\rho_0$)/T$^2$ as a function of T$^3$ which shows a negative slope (inset (b)) indicating coefficient of T$^5$ term in resistivity is negative \cite{sch67, sch68}. The estimated mean free path $l$ of the conduction electrons is estimated in the framework of free electron model as $l (\times 10^{-9} m) = (r_s/a_0)^2 *9.2 /\rho_{\mu}$, where $r_s$ is the radius of sphere whose volume is equal to the volume per conduction electron, a$_0$ is he Bohr atomic radius and $\rho_{\mu}$ is the residual resistivity expressed in $\mu \Omega$-cm \cite{ash01}. In Ti$_{0.6}$V$_{0.4}$ alloy, the mean free path estimated to be about 0.3 x 10$^{-9}$~m which is of the order of unit cell.  
In case of large residual resistivity values such as 99 x 10$^{-8}$~$\Omega$m, the estimation of the mean free path is not true as Matthiessens rule is not valid \cite {moo73}. In such cases, the superconducting property can be used to estimate the mean free path \cite{hak67}. In the dirty limit superconductor, the coherence length is limited by the mean free path and hence $l = \xi(0)$. Then the mean free path in this alloy is expected to be about 4.5 x 10$^{-9}$~m. According to Mooij, short mean free path resulting from various scattering mechanism such as $s-d$ interaction, disorder, magnetic interaction etc., is the reason for the small temperature variation of resistivity \cite{moo73}.


The figure 7 shows the temperature dependence of the heat capacity C for Ti$_{0.6}$V$_{0.4}$ in the temperature range 20-220~K. The fitted line represents the expression $\gamma$T + D(T/$\theta_D$), where $\gamma$ is the Sommerfeld coefficient of electronic Specific heat and D(T/$\theta_D$) is the Debye specific heat for lattice. A good fitting to the experimental heat capacity curve is obtained only in the high temperature region (150-220~K). The value of $\gamma$ and $\theta_D$ obtained as the fitting parameters are about 8.06(1) x 10$^{-3}$~J/mol K$^2$ and 322.5(4)~K respectively. The fitted $\gamma$T + D(T/$\theta_D$) curve deviates from the experimental curve below 105~K (inset to the Fig. 7.). The figure 8 shows the C/T plotted as a function of T$^2$ in the temperature range 7-45~K. A nonlinearity is observed in C/T with respect to T$^2$. Such nonlinearity as well as the deviation from $\gamma$T + D(T/$\theta_D$) might originate from a finite temperature dependence of the Debye temperature $\theta_D$ in a bcc structure \cite{fin39, cho58}. Normally, the $\theta_D$ increases with decrease in temperature below 30~K, especially in case of bcc systems the values at low temperature increases more than that at room temperature \cite{cho58, gop66}. The variation of $\theta_D$ for elemental bcc vanadium in the temperature range 4.2~K to 300~K estimated by measuring elastic constants is about 403.8~K to 392.1~K which is about 3~\% \cite{bol71}. In Ti$_{0.6}$V$_{0.4}$ alloy we have observed that $\theta_D$ estimated from the low temperature fit is about 258~K which is quite low when compared to that estimated at high temperatures. Hence, we believe that the origin of the difference between the experimentally observed heat capacity and the heat capacity estimated using $\gamma$T + D(T/$\theta_D$) is not due to the temperature variation of the $\theta_D$.  However, we have done the thermal expansion measurement to estimate the Debye temperature \cite{gop66, say98} as function of temperature \cite{unp12}. We have observed that the variation in $\theta_D$ with temperature in the range 30-300~K is about {7.5~\%}. Hence, the deviation observed in heat capacity of Ti$_{0.6}$V$_{0.4}$ from $\gamma$T + D(T/$\theta_D$) below 100~K is not due to variation in $\theta_D$ with temperature. The difference may also arise from the acoustic phonons which show peak structures in the phonon dispersion curves. These modes act as a Einstein oscillators and hence, we have checked for contributions from Einstein modes obtained from the phonon dispersion curve reported in literature for Vanadium \cite{col70}. We have observed that for any value of Einstein temperature $\theta_E$, the value of $\theta_D$ diverges during fitting. Hence, it appears that the contribution to the heat capacity due to the Einstein modes is negligible for the present system. By comparing our results with the literature we found that $\Delta C$ is observed to follow a temperature dependence such as DT$^3$ln(T) \cite{ike82, koh08}. This is verified by the presence of linearity in $\Delta C/T^3$ as a function of lnT (inset a to Fig 8.).


Whenever, T$^2$ dependence in resistivity is observed, generally the Kadowaki-Woods relation is checked for its validity. The inset (b) to the figure 8 shows the Kadowaki Woods plot for Ti$_{0.6}$V$_{0.4}$ along with several other spin fluctuations and heavy fermion systems. Surprisingly, the Kadowaki-Woods relation is observed to be valid in our system. Kadowaki and Woods have empirically shown that the coefficient $A$ of T$^2$ term in resistivity scales with the square of the coefficient $\gamma$ of electronic specific heat for heavy Fermions and spin-fluctuation systems \cite{kad86}. The proportionality constant (Kadowaki-Woods' ratio, A/$\gamma^2$) is found to be about 1 x 10$^{-5} \mu \Omega$-cm (mol K/mJ)$^2$ \cite{kad86}. The coefficient A of the T$^2$ term in resistivity for the present Ti$_{0.6}$V$_{0.4}$ is estimated to be about 5.00(1) x 10$^{-4}~\mu \Omega$-cm /K2 (inset to Fig. 6.). The ratio (A/$\gamma^2$) for Ti$_{0.6}$V$_{0.4}$ is about 0.60(5) x 10$^{-5}$ $\mu \Omega$-cm (mol K/mJ)$^2$ which is close to the proportionality constant obtained by Kadowaki-Woods'.

We summarize below the important observations made in the present set of experimental results on the normal state of Ti$_{0.6}$V$_{0.4}$ alloy.

1. The temperature dependence of dc susceptibility shows the characteristic "temperature induced magnetic moment" and follows $\chi$(T) =$\chi$(0) -bT$^2$ ln(T/T$^*$).

2. The magnetization is linear in H.

3. Resistivity shows a T$^2$ dependence at low temperatures.

4. The coefficient of T$^5$ term in resistivity is negative.

5. There is an enhancement of Sommerfeld coefficient of electronic specific heat $\gamma$.

6. The excess heat capacity ($\Delta$ C = C - $\gamma$T - D(T/$\theta_D$)) follows a DT$^3$ln(T) temperature dependence.

7. Kadowaki-Woods' relation is valid in this system. 

All these features together have been observed only in Pauli enhanced paramagnets such as RCo$_2$(R = Y, Sc, Lu, etc.,) \cite{ike84, hau98, gra95}, etc., where the physical properties have been explained in terms of the existence of itinerant spin fluctuations. We argue that itinerant spin fluctuations may be present in the Ti$_{0.6}$V$_{0.4}$ alloys as well.

\subsection{Itinerant spin fluctuation model and its significance in Ti$_{0.6}$V$_{0.4}$}

We now present a model based on the existence of itinerant spin fluctuations in the Ti$_{0.6}$V$_{0.4}$ alloy and show that the observed superconducting properties of the alloy can be explained in this picture. In our understanding, the characteristic temperature T$^*$ = 90~K observed in resistivity is the spin fluctuation temperature T$_{SF}$ \cite{gra95}. It has been shown by Frings and Franse that the T$_{SF}$ may be identified as the temperature at which the second derivative of the susceptibility with respect to temperature is zero \cite{fri85}. In the present Ti$_{0.6}$V$_{0.4}$ alloy, the second derivative of susceptibility goes to zero  at T$_{SF} \sim$ 120~K. The T$_{SF}$ estimated from the straight line fit to data $\Delta C/T^3$ v/s lnT is about 53~K. However, T$_{SF}$ estimated by resistivity, dc susceptibility and heat capacity are in the same order of magnitude. Variation of T$_{SF}$ among various experiments to about 2T$_{SF}$ is not uncommon, and is often found in literature \cite{fri85}. 

The presence of spin fluctuations in a material can be characterized through the Stoner enhancement factor \cite{von82}. The Stoner enhancement factor S is defined as $\chi$(0)/$\chi_0$ where $\chi$(0) is the experimentally observed susceptibility and $\chi_0$ is the susceptibility due to density of states \cite{von82, can07}. The $\chi_0$ (dimensionless) is given as (3/2)$\mu_0 \mu_{B}^{2}n/\epsilon_F$ = 3$\mu_0 \mu_{B}^{2} \gamma /\pi^2 k_{B}^{2}$ where $\mu_0$ is the permittivity of the free space, $\mu_B$ is the Bohr magneton, $n$ is the free electron density, $\epsilon$ is the Fermi energy, and $k_{B}$ is the Boltzmann constant \cite{ash01}.  The value of S for strong spin fluctuating systems like Pd or Pt is about 6 \cite{von82, rie79} where as the value of S for superconducting Nb or V is about 2 \cite{von82, rie79}. The value of $\chi_0$ for Ti$_{0.6}$V$_{0.4}$  is about 1.57 x 10$^{-4}$. Then for Ti$_{0.6}$V$_{0.4}$ the value of S = 2.5, where experimentally observed $\chi$(0) = 3.92 x 10$^{-4}$ (dimensionless). This value of S is similar to that of V and Nb and is much less than that of metals like Pt and Pd \cite{von82, rie79}. This is also similar to the value of S for Mo$_3$Sb$_7$ system \cite{can07}. 

The Sommerfeld coefficient of electronic heat capacity $\gamma$ enhances due to the presence of spin fluctuations as $\gamma$ = $\gamma_0 (1 + \lambda_{ep} + \lambda_{SF})$ where  $\gamma_0 (1 + \lambda_{ep})$ is the Sommerfeld coefficient of electronic heat capacity without spin fluctuations. Our results suggest that the spin fluctuations are absent temperatures more than about 100~K. Our fit to the heat capacity at temperatures above 100~K provided a value of $\gamma$ to be about $\gamma_{HT}$ = 8.06 x 10$^{-3}$ J/molK$^2$ whereas the low temperature fit provided a value of $\gamma$ to be about $\gamma_{LT} = $9.46 x 10$^{-3}$ J/molK$^2$. Then the value of $\lambda_{SF}$ can be estimated as $(\gamma_{LT}/\gamma_{HT})-1$  which turns out to be about 0.17.

The two major effects of the itinerant spin fluctuations on the superconducting properties in the Ti$_{0.6}$V$_{0.4}$ alloy are: (i)	 The enhancement of superconducting transition temperature in Ti$_{0.6}$V$_{0.4}$ with respect to Vanadium, and (ii)	The origin of the fluctuation conductivity above T$_c$ and above H$_{c2}$.

These effects are discussed below:

(i) The enhancement of superconducting transition temperature in Ti$_{0.6}$V$_{0.4}$ with respect to Vanadium:

The itinerant spin-fluctuations are reported to be responsible for the absence of superconductivity in heavy transition elements such as Palladium and Platinum \cite{don67, ber66, von82}. The same formalism was used to suggest that the superconducting transition temperature T$_c$ of elements such as Nb and V is limited by spin-fluctuations \cite{rie79, wei05, von82}. The spin fluctuation coupling constant $\lambda_{SF}$ in V is estimated to be about 0.34 \cite{von82}. Then, when non magnetic Ti is alloyed with Vanadium, there is a reduction in spin fluctuation, hence, T$_c$ increases in spite of decrease in the free electron density. At very high content of Ti, the T$_c$ again starts decreasing when the effect of loss of free electron density wins over the effect of reduction of spin fluctuations. The value of 2-3 times T$_c$ falls in the range of the superconducting transition temperatures of Ti-V alloys calculated by estimating the electron-phonon coupling constant $\lambda_{ep}$ from band structure calculation (which is about 12-20~K) \cite{pic87}. Hence, we can estimate the $\lambda_{SF}$ for Ti$_{0.6}$V$_{0.4}$ by considering the suppression of T$_c$ from the theoretical limit (in absence of spin fluctuations) of about 14.3~K to the experimentally observed value of 7.15~K (in presence of spin fluctuations) as described below. Daams et al have shown that the properties of superconductors with spin fluctuations can be scaled to that of a superconductor without spin fluctuation by introducing the renormalized parameters $\lambda_{eff}$ and $\mu^*_{eff}$ \cite{daa81}. In this formalism, the superconducting transition temperature is given by the modified McMillan formula \cite{can07} as

\begin{equation}	
T_C = (\theta/1.45)*exp\left(\frac{-1.04(1+\lambda_{eff})}{\lambda_{eff} - \mu^*_{eff}(1+0.62\lambda_{eff})}\right), 
\end{equation}

where $\lambda_{eff} = \lambda_{ep}$(1+$\lambda_{SF})^{-1}$ and $\mu^*_{eff} = (\mu^*+\lambda_{SF})(1+\lambda_{SF})^{-1}$. Here, $\mu^*$ is the Coulomb coupling constant and $\lambda_{SF}$ is the electron-spin fluctuation coupling constant.  

Using this formalism, the $\lambda_{ep}$ and $\lambda_{SF}$ for Ti$_{0.6}$V$_{0.4}$ are estimated to be about 0.85 and 0.081 respectively. The value of $\lambda_{SF}$ estimated in this way is in agreement with that estimated from the enhancement of $\gamma$ at low temperatures. The value of $\mu^*$ is taken to be 0.1 \cite{von82}. The other parameters $\lambda_{eff}$ and $\mu^*_{eff}$ were estimated to be 0.79 and 0.17 respectively. These values are used to analyze temperature dependence of H$_{c2}$ on the basis of formalism by Orlando et al \cite{orl79}. The solid line in the Fig. 3b., shows the temperature dependence of H$_{c2}$ estimated in this manner \cite{orl79}. The spin-orbit interaction parameter $\lambda_{SO}$ is estimated to be about 0.3.

(ii) The origin of the fluctuation conductivity above T$_c$ and above H$_{c2}$:

When Vanadium is diluted with the Titanium without changing the bcc structure, the spin fluctuations become inhomogeneous in space due to the spatial disorder in Vanadium sublattice. In such a case electron spin-fluctuation coupling constant $\lambda_{SF}$ is expected to vary in space over the sample volume. This is similar to the disordered Kondo lattice with spatial distribution of Kondo temperatures T$_K$'s \cite{ste01} leading to inhomogeneous short range conduction electron polarization which in turn results in the itinerant spin fluctuations. Since, the mobility of the conduction electrons are large, the life time of these spin fluctuations decreases leading to the small spin fluctuation coupling constant $\lambda_{SF}$ of about 0.081 in the Ti$_{0.6}$V$_{0.4}$ alloy. Our conjecture about the itinerant nature of the spin fluctuations is also supported by the fact that there is no significant magnetic field dependence of physical properties above 20~K, where the signature of the existence of spin fluctuations are observed in the Ti$_{0.6}$V$_{0.4}$ alloy \cite{kus12}. Since the itinerant spin fluctuations are weakly dependent on the magnetic field, the $\sqrt(H)$ dependence of the (negative) magneto-resistance is absent in Ti$_{0.6}$V$_{0.4}$.

\section {Summary and conclusion}
The temperature and magnetic field dependent dc magnetic susceptibility and electrical resistivity and the zero field heat capacity of Ti$_{0.6}$V$_{0.4}$ have been analyzed in a quantitative manner. These results provide some experimental evidences of the presence of spin-fluctuations in this alloy system. The Kadowaki-Woods scaling is shown to be valid for the Ti$_{0.6}$V$_{0.4}$ alloy. We have argued that the spin fluctuations present in Ti$_{0.6}$V$_{0.4}$ alloy are itinerant in nature. The onset of superconductivity is suppressed in Ti$_{0.6}$V$_{0.4}$ from its expected theoretical limit due to the presence of such spin-fluctuations while the distribution of $\lambda_{SF}$ induces the superconducting fluctuation above T$_c$. There are some indications based on the present study of the existence of preformed Cooper-pairs in the temperature range well above the superconducting transition temperature. The present study suggests that the interesting correlations between spin-fluctuations and superconductivity may not necessarily be the properties of only certain classes of exotic compounds.  Relatively simple alloy systems like Ti$_{0.6}$V$_{0.4}$ shows such interesting correlations. A complete understanding of the normal state and superconducting properties of this technologically important alloy system Ti-V would help in tuning its properties for suitable technological applications.

\section {Acknowledgments}
Authors would like to thank Ms. Parul Arora for the heat capacity measurements, and Mr. V. K. Sharma for magnetization measurements using SQUID magnetometer. Dr. S. K. Deb, Head, ISUD, RRCAT and Dr. P. D. Gupta, Director, RRCAT are acknowledged for their support in the experiments done at Indus-2 Synchrotron Radiation Source at RRCAT, Indore.

\newpage

\noindent {\large Figure Captions :}\\

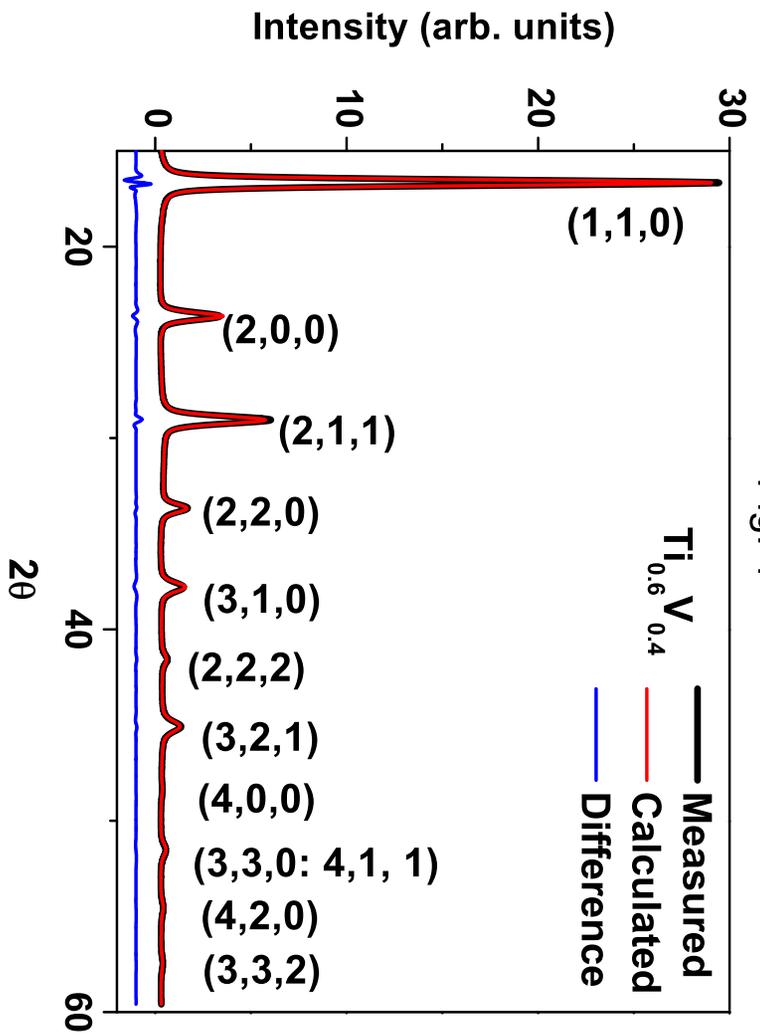
\noindent Figure 1. (color online) The x-ray diffraction pattern of Ti$_{0.6}$V$_{0.4}$ in the 2$\theta$ range 15-60~$^o$. The Rietveld analysis (red solid line) shows that the sample is formed in bcc structure with lattice parameter 0.3188~nm.

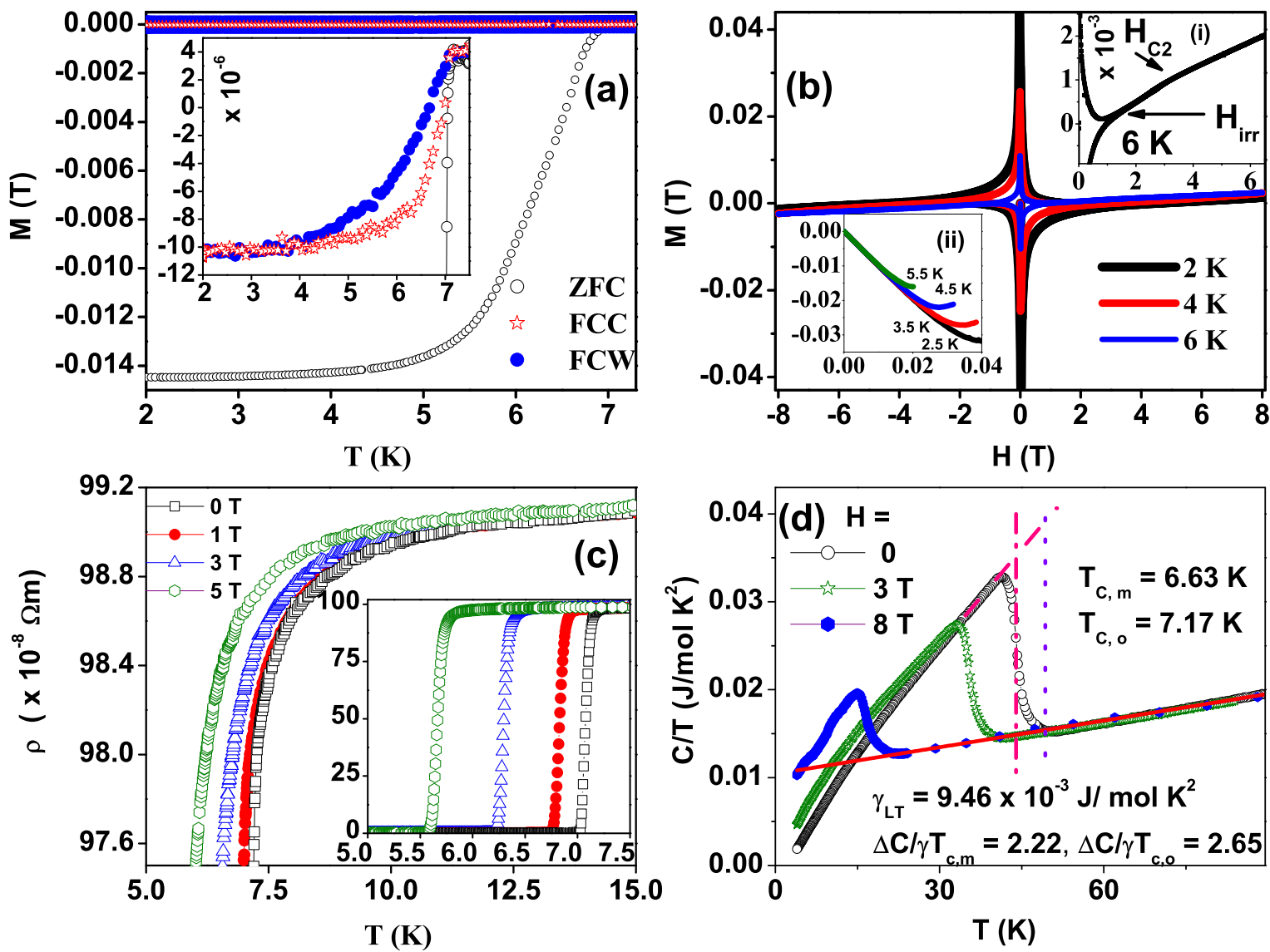
\noindent Figure 2. (color online) (a) Temperature dependence of magnetization below 7.2~K for 10~mT in ZFC, FCC and FCW mode. There is a large difference between ZFC and FCC/FCW values indicating strong pinning of the vortices.

(b) Magnetization as a function of magnetic field at various temperatures below T$_c$ for Ti$_{60}$V$_{40}$. The inset (i) shows the magnetization as a function of magnetic field near H$_{c2}$, where H$_{c2}$ and H$_{irr}$ are marked by arrows. The inset (ii) shows the low field magnetization data which is used to estimate H$_{c1}$.  

(c) Temperature dependence of resistivity of Ti$_{0.6}$V$_{0.4}$ in the range 6-15~K at various magnetic fields up to 5~T. A rounding-off behavior is observed above T$_c$. The inset shows the temperature dependence of resistivity across the superconducting transition temperature.

(d) Temperature dependence of heat capacity of Ti$_{0.6}$V$_{0.4}$ in zero, 3~T and 8~T magnetic field. The symbols are the experimental data points. The red straight line is fit to the normal state data. The dotted straight line is the guide to eye which is used to estimate the jump in heat capacity at T$_{c,o}$ = 7.17~K whereas the dashed line is the guide to eye which is used to estimate the jump in heat capacity at T$_{c,m}$ = 6.63~K.

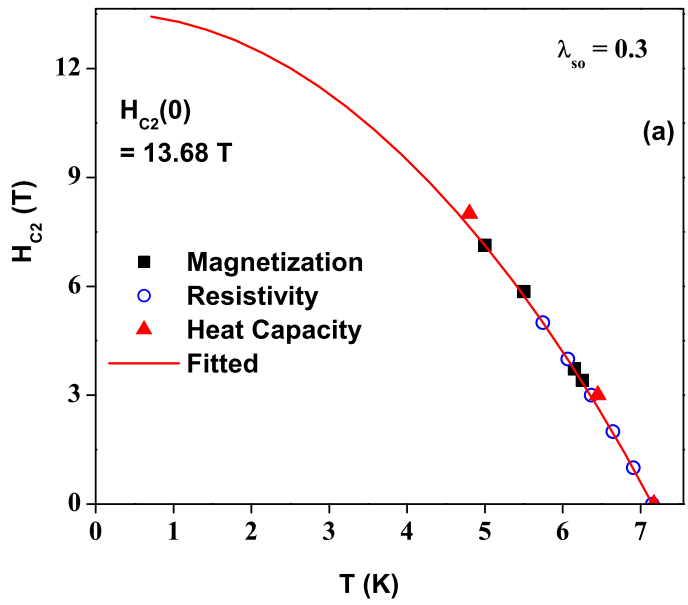
\noindent Figure 3. (color online) (a) The temperature dependence of H$_{c2}$ for Ti$_{0.6}$V$_{0.4}$. Symbols represent the experimental data points, solid line is fit to the data using formalism from Orlando et. al.\cite{orl79} (b) The temperature dependence of H$_{c1}$ for Ti$_{0.6}$V$_{0.4}$. Symbols represent the experimental data points, solid line is fit to the data using the formula H$_{c1}$(T) = H$_{c1}$(0) [1-(T/T$_c$)$^2$].

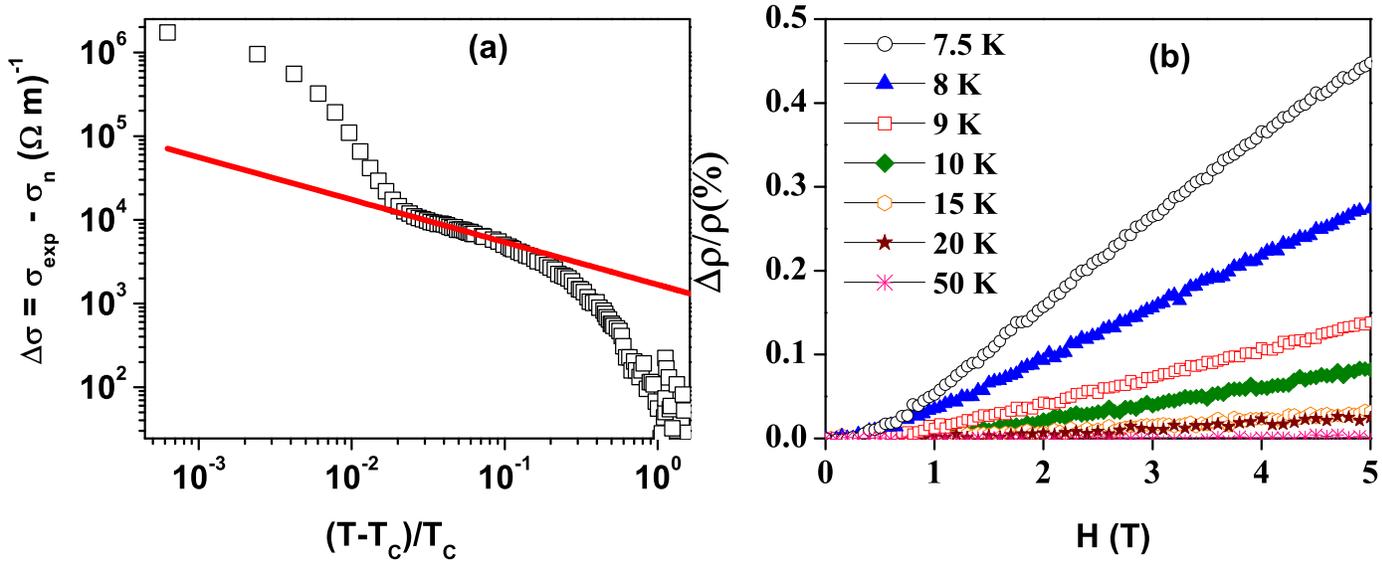
\noindent Figure 4. (color online) (a)The $\Delta \sigma$ = $\sigma_{exp}$- $\sigma_n$  as a function of (T-T$_c$)/T$_c$ in a log-log scale for Ti$_{0.6}$V$_{0.4}$.The straight line fit to the data shows that $\Delta \sigma$ follows the Eq. 1., with $\alpha \sim$ 0.5 indicating the existence of 3D fluctuations.

(b) Magneto-resistance as a function of magnetic field up to 5~T above T$_c$ = 7.06~K for Ti$_{0.6}$V$_{0.4}$. The magneto-resistance is about 0.45~\% at 7.5~K and 5~T. For temperatures above 20~K, the magneto resistance is zero.

\noindent Figure 5. (color online) (a)Temperature dependence of dc susceptibility ($\chi$ = M/H) of the Ti$_{0.6}$V$_{0.4}$ alloy in the range 10-300~K in 1~T magnetic field. As the temperature decreases below at 300~K, the $\chi$(T) decreases down to 20~K before showing a low temperature upturn. The red solid line shows the fit to the experimental data using Eq. 2. The inset shows the normalized $\chi$ with respect to $\chi$(30~K) measured at 1~T (black open circles) as well as $\chi$ estimated from high field M v/s H data (Fig. 5(b)) at various temperatures (Red filled squares). 

(b) The plot of magnetization M as a function of $\mu_0$H at 10~K, 30~K and 300~K. The solid black line is the field dependence of magnetization measured at 10~K using vibrating sample magnetometer whereas the black open squares are the M v/s H data measured using SQUID magnetometer. The M is linear in H up to 8~T. The inset shows the plot of difference in magnetization between 300~K and various temperatures to highlight the variation in field dependecne of magnetization with temperature.
 
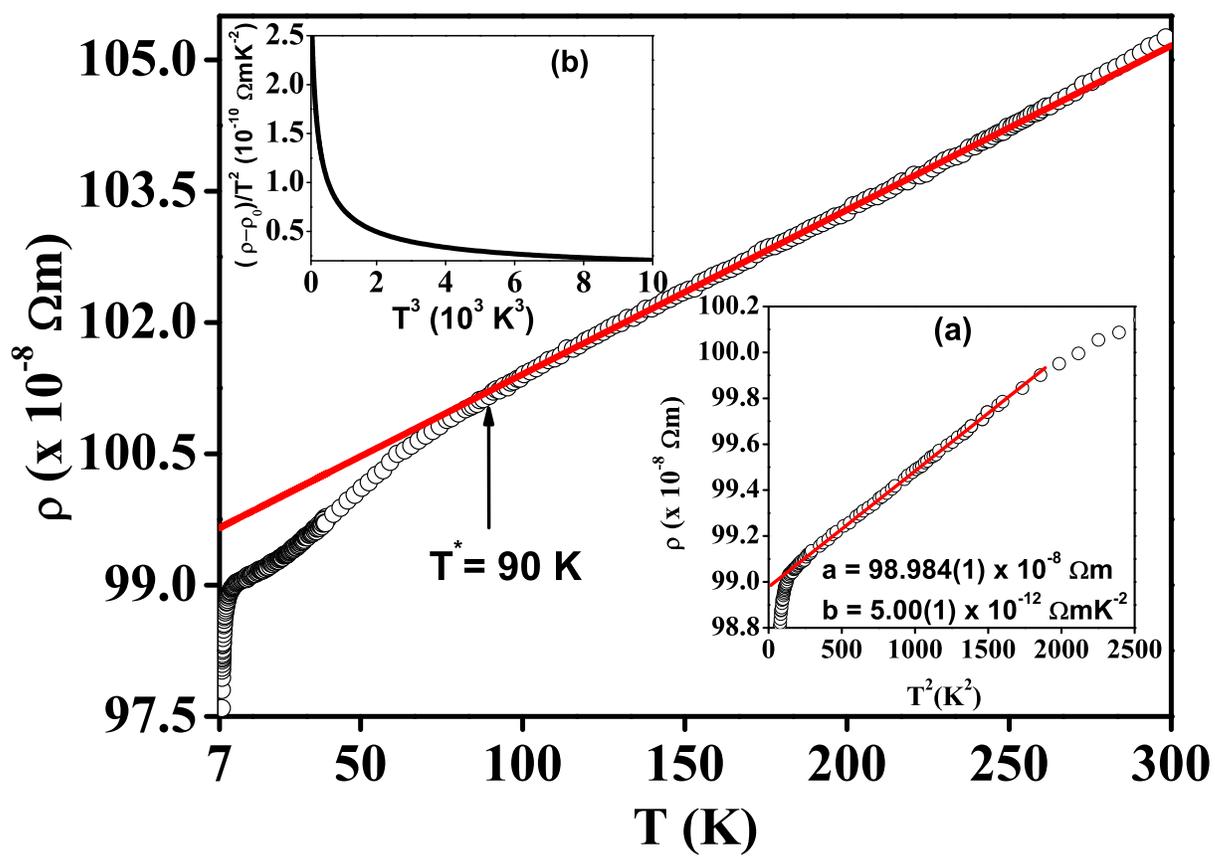
\noindent Figure 6. (color online) The temperature dependence of resistivity of Ti$_{0.6}$V$_{0.4}$ at zero magnetic field in the temperature range 5-300~K. Resistivity is linear (red dotted line) in the temperature interval of 300-100~K. Resistivity drops sharply below T$^*$ = 90~K before it drops to zero at 7~K. Inset (a) shows the resistivity of Ti$_{0.6}$V$_{0.4}$ as a function of square of temperature in the range T = 20~K to 50~K. A linear fit to the data indicates $\rho$ is proportional to AT$^2$ at low temperatures. Inset (b) shows the plot of $\rho -\rho_0$/T$^2$ as function of T$^3$ in the temperature range 10-100~K. The negative slope of the curve is an indicative of negative coefficient of T$^5$ term in resistivity.

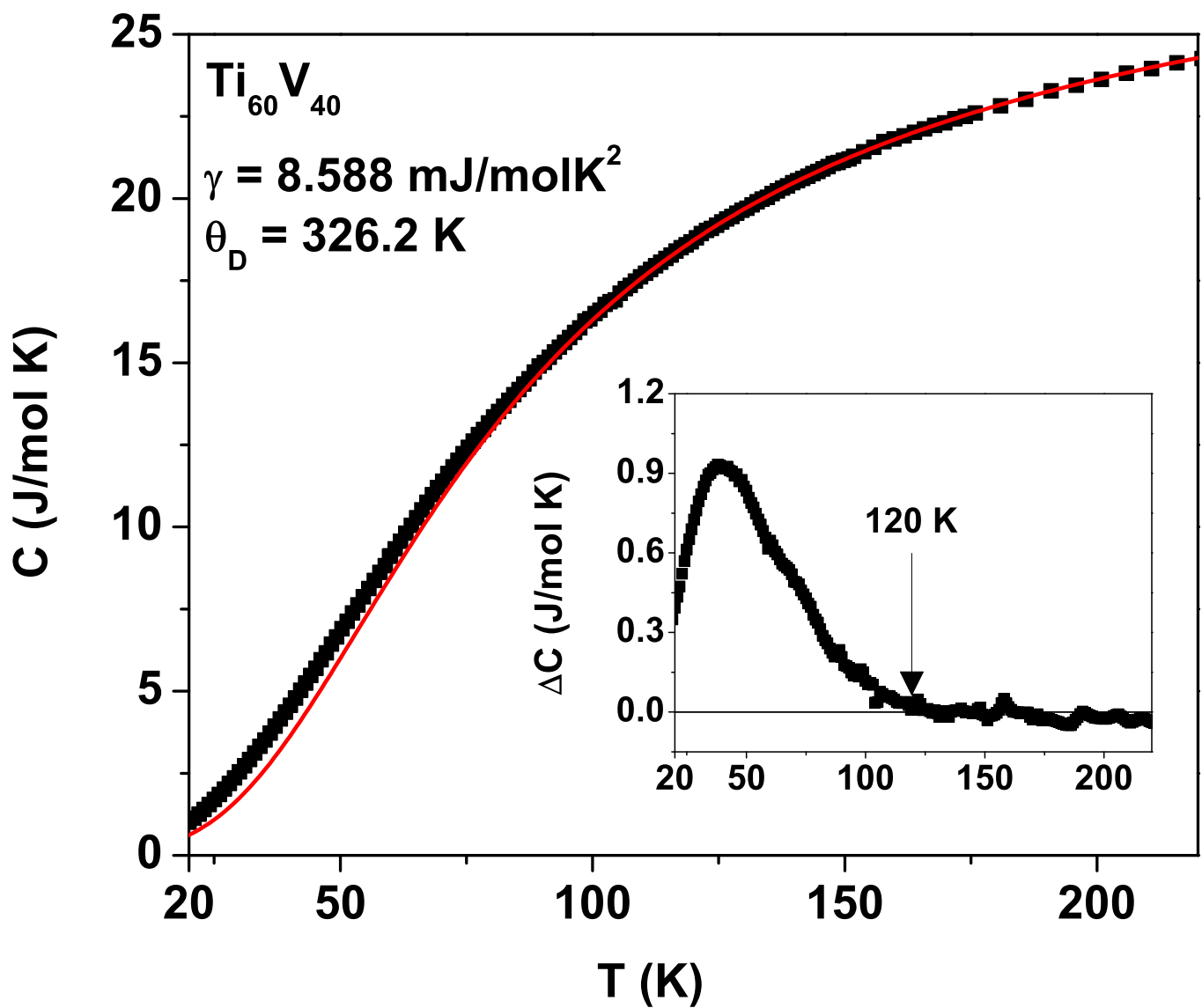
\noindent Figure 7. (color online) The temperature dependence of heat capacity in the range 20-220~K. The solid squares are the experimental data points. The red solid line is fit to the data in the range 150 - 220~K using $\gamma$T + D(T/$\theta_D$), where D(T/$\theta_D$) is the Debye specific heat for lattice. The deviations are observed below 105~K (inset to the figure).

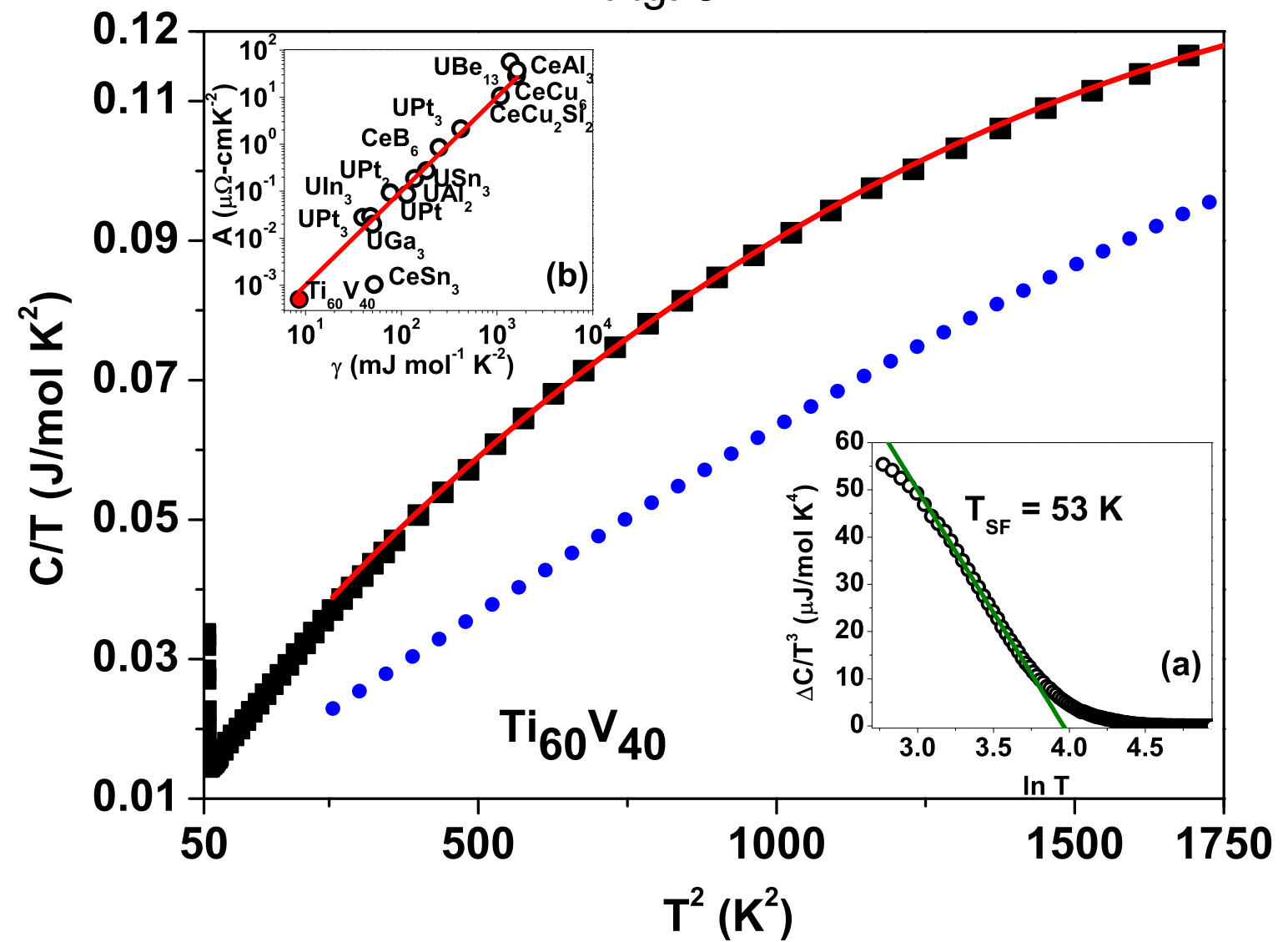
\noindent Figure 8. (color online) The C/T for Ti$_{0.6}$V$_{0.4}$ as function of T$^2$ below 45~K.  The deviation from linearity is observed in this temperature range.  The dotted blue line is the heat capacity data estimated by considering $\gamma$ = 9.46(1) x 10$^{-3}$~J/molK$^2$ and $\theta_D$ = 322.5(4)~K. The solid red line is the heat capacity estimated after considering spin fluctuations along with the lattice and electronic contributions. The inset (a) shows the linear dependence of $\Delta$ C/T$^3$ on lnT at low temperatures. The inset (b) shows the validity of Kadowaki-Woods plot for Ti$_{0.6}$V$_{0.4}$.

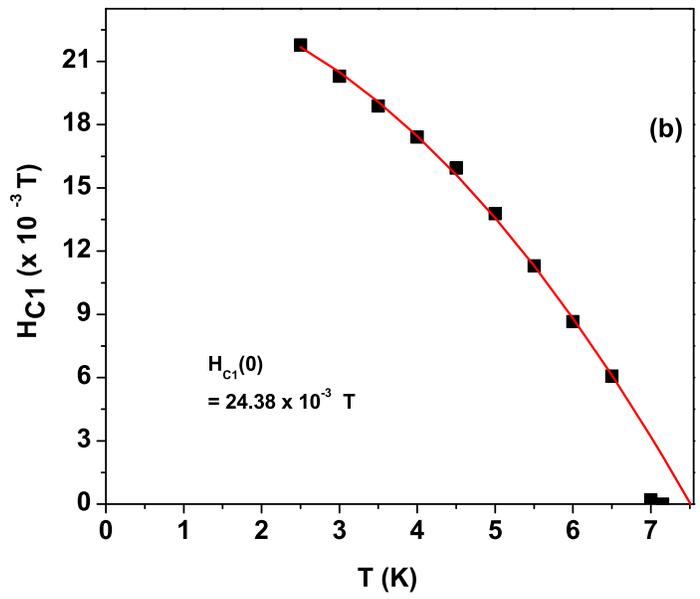
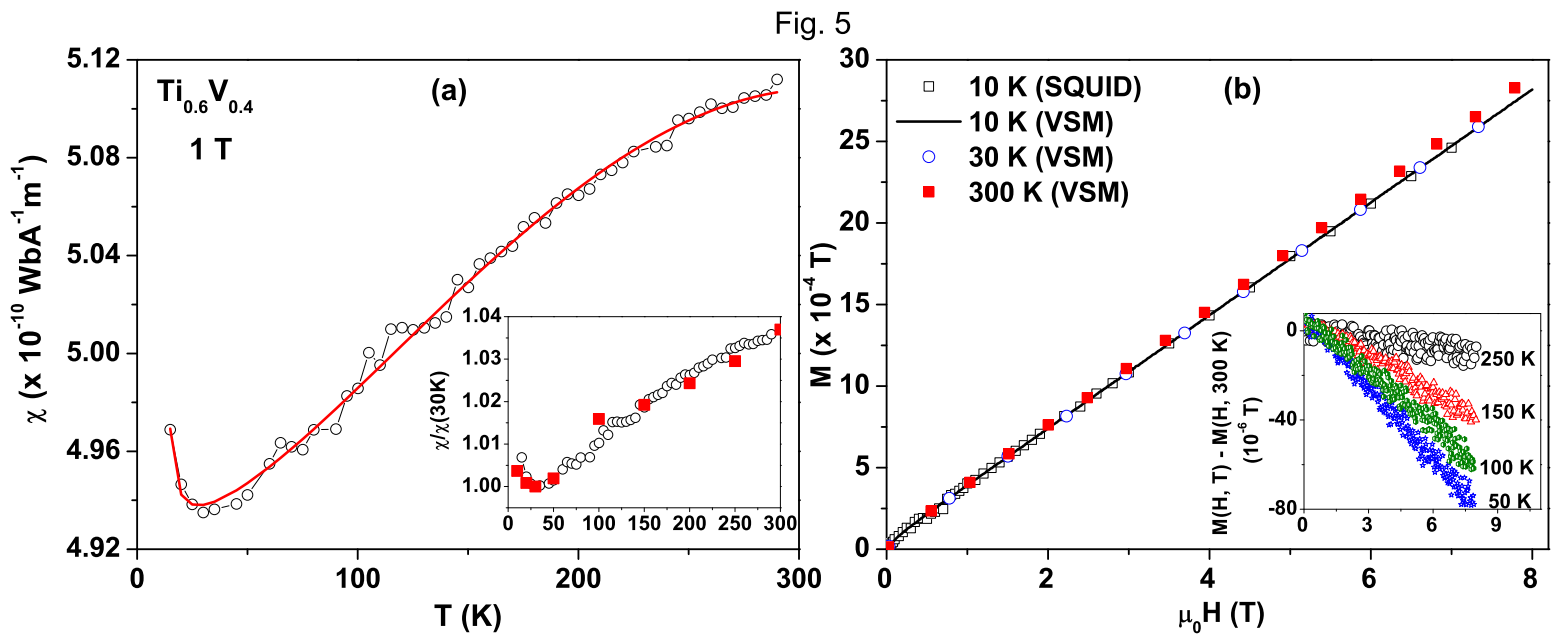
\end{document}